\documentclass[aps,prb,twocolumn,epsfig,floatfix,twoside,a4paper,showpacs]{revtex4}

\usepackage{epsfig}

\begin{document} 
\input epsf 
\title{Effect of concentration on 
the thermodynamics of sodium chloride aqueous solutions 
in the supercooled regime} 
\author{
D.~Corradini, P.~Gallo\footnote[1]{Author to whom correspondence
should be addressed;  e-mail: gallop@fis.uniroma3.it} and M.~Rovere }
\affiliation{Dipartimento di Fisica, Universit\`a ``Roma Tre'' \\ 
Via della Vasca Navale 84, I-00146 Roma, Italy\\}

\begin{abstract}
\noindent Molecular Dynamics simulations are performed on two sodium
chloride solutions in TIP4P  water with concentrations $c=1.36\, mol/kg$
and $c=2.10\, mol/kg$ upon supercooling.  The isotherms and isochores
planes are calculated.   The temperature of maximum density line and
the limit of mechanical stability  line are obtained from the
analysis of the thermodynamic planes.  The comparison 
of the results shows that for 
densities well above the limit of mechanical stability, the isotherms 
and isochores  of the sodium chloride aqueous solution shift to lower pressures 
upon increasing  concentration while
the limit of mechanical stability is very  similar to that of bulk water
for both concentrations.
We also find that the temperature of maximum density line shifts to lower pressures 
and temperatures  upon increasing concentration. Indications of the presence of
a liquid-liquid coexistence are found for both concentrations. 

\end{abstract}



\maketitle

\section{Introduction}\label{intro}

The properties of aqueous ionic solutions besides being 
of undoubtful importance in chemical physics\cite{Angell} and
electrochemistry~\cite{Anderko},  are relevant in many other fields
of science including biology and biophysics~\cite{Leckband}, 
geophysics~\cite{Sherman}, and even atmospheric modeling.~\cite{Archer}
In the supercooled region, thermodynamic properties of
solutions are also of interest for the cryopreservation of
organs and food.~\cite{Urrutia,pablobook,franksbook}
From a more fundamental point of view 
an improved understanding of the
thermodynamics of these systems upon supercooling, can help to shed light
on the open questions on bulk liquid water.~\cite{Angell2} 

It is well known that water
presents, in the supercooled region, peculiar thermodynamic
behavior.~\cite{phystoday,Debenedetti,Genegroup,Angell3,Angell4} In
particular, the most striking effects are the existence of a
temperature of maximum density (TMD) line  and the divergence of the
isothermal compressibility $K_T$, of the isobaric specific heat $c_P$
and of the coefficient of thermal expansion $\alpha_P$.  The origin of
this anomalous behavior is still a matter of large
interest and debate in the literature.~\cite{Debenedetti} 
Several
theoretical\cite{Roberts,Cervantes,Widom,Franzese,Kumar2,Truskett,Grande,Jefferey} 
and computer
simulation\cite{Gene,Poole,Paschek, Paschek2, Tanaka, Brovchenko, Vallauri, Mossa, Sciortino} 
papers have shown
the presence in the supercooled region of water of a
liquid-liquid (LL) critical point. Experimental signatures 
of this critical point have been also found.~\cite{Mishima}
The second critical point of water
would be the end point of the coexistence line between a low density
liquid (LDL) and a high density liquid (HDL). In this
framework, the anomalous properties of water
arise as a consequence of the presence of the LL
critical point. Furthermore in this picture, the limit of mechanical stability
(LMS) is non-reentrant and the TMD line is knee-shaped and avoids to cross the LMS line.

Aqueous ionic solutions have been
extensively studied at ambient temperature.
Their structural properties are the main focus of
most papers, with particular emphasis
on the hydration structure.~\cite{leberman,degreve,koneshan,bruni,chowduri,hribar,chandra,jardon,patra,botti,du,mancinelli,panag}
Many studies in the supercooled regime deal with the glass transition phenomenon
(see Ref.~\onlinecite{Angell} and references therein) 
while the detailed comparison of
the thermodynamic behavior of the aqueous solutions with respect to bulk water
in the mild supercooled regime still lacks a thorough investigation.
Calorimetric experiments
have shown that from low to moderate concentration of ions 
several thermodynamic properties of aqueous solutions are dominated by the
solvent.\cite{Archer,Archer2} 

In this paper, we present a Molecular Dynamics (MD) simulation study
of the thermodynamics of two sodium chloride aqueous solutions,
in the following denoted also as NaCl(aq), in the supercooled
regime. This work is an extension of a previous study performed on
bulk water and on a NaCl(aq) solution with low salt concentration.~\cite{mio} 

The concentrations of salt in the solutions studied in the present work
are $c=1.36\, mol/kg$ and $c=2.10\, mol/kg$.  For both systems we study the
isotherms in the $P-\rho$ plane and the isochores in the $P-T$
plane. The analysis of those thermodynamic planes leads to the
determination of the LMS and TMD lines. Moreover we present the trend
of the potential energy as a function of density, at a low temperature.
We will compare the results of the present simulations with results on
bulk water and $c=0.67\, mol/kg$ NaCl(aq)
studied in our previous work.~\cite{mio} We also 
perform a comparison of the results with what found for water confined in
a hydrophobic environment of soft
spheres.~\cite{gallorovere}  

The paper is organized as follows. In
Sec.~\ref{model} we explain the details of the model and the
computer simulation setup. In Sec.~\ref{thermo} we show and discuss
the  thermodynamic behavior. Conclusions are drawn in Sec.~\ref{conclu}.

\section{Model and simulation details}\label{model}

\begin{figure}[t] \centerline{\psfig{file=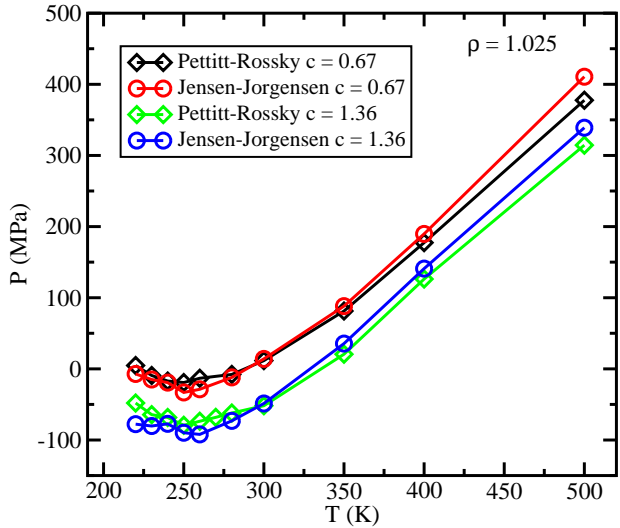,
width=0.5\textwidth}}
\caption{Isochores
for $c=0.67\, mol/kg$ NaCl(aq)~\cite{mio} and  
for $c=1.36\, mol/kg$ NaCl(aq) for
$\rho=1.025$ $g/cm^3$ 
for two different force
fields~\cite{rossky,JJ} (color online).}
\label{fig:1}
\end{figure}

\begin{figure}[t] \centerline{\psfig{file=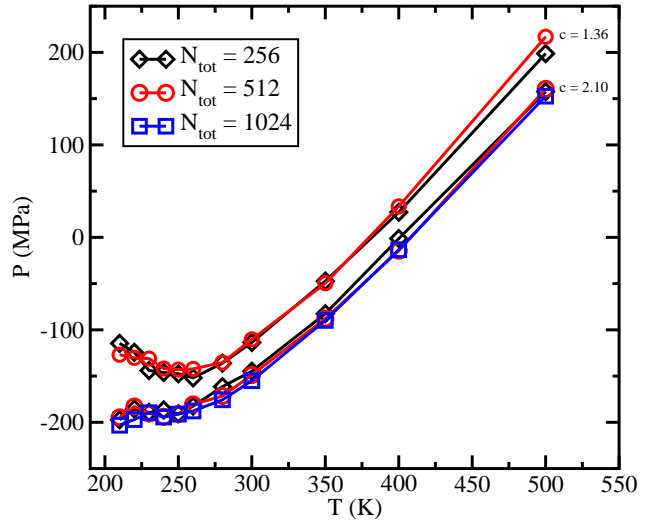,
width=0.5\textwidth}}
\caption{Isochores
for $c=1.36\, mol/kg$ NaCl(aq) and  
for $c=2.10\, mol/kg$ NaCl(aq) for
$\rho=0.98$ $g/cm^3$ 
for three different sistem sizes (color online).}
\label{fig:2}
\end{figure}

Two aqueous sodium chloride solutions with concentrations,  given in moles of
solute per mass of solvent, $c=1.36\, mol/kg$  and $c=2.10\, mol/kg$ 
are simulated by means of MD technique.  
In the case of the $c=1.36\, mol/kg$ solution the system is composed by
244 water molecules, 6 $Na^+$ ions and 6 $Cl^-$ ions, while in the
case of the $c=2.10\, mol/kg$ solution, it is composed by 238 water
molecules, 9 $Na^+$ ions and 9 $Cl^-$ ions.  

The particles interact
via the sum of coulombic and Lennard-Jones (LJ) potentials. 
The analytical expression of the interaction potential is given by
\begin{equation} U_{ij}(r)=\frac{q_i q_j}{r}+4\epsilon_{ij}
\left[\left(\frac{\sigma_{ij}}
{r}\right)^{12}-\left(\frac{\sigma_{ij}}{r}\right)^{6}\right]\, 
\end{equation} 
where $q$ is the electric charge 
and  $\epsilon_{ij}$ and $\sigma_{ij}$ are LJ parameters.
Water molecules are
modeled employing the TIP4P potential.~\cite{tip4p} 
Ion-ion and ion-water LJ parameters are derived from Pettit and
Rossky~\cite{rossky}  parameters for the Huggins-Mayer potential, via
the reparametrization made by  Koneshan and Rasaiah~\cite{koneshan}
for LJ potential. The ion-water and ion-ion 
LJ interaction parameters are summarized in
Table~\ref{tab:tab1}.

\begin{table}[htbp]
\caption{Ion-water and ion-ion  LJ interaction
parameters.}
\begin{center}
\begin{ruledtabular}
\begin{tabular}{lcc} Atom pair & $\epsilon\, (\mbox{kJ/mol})$ &
$\sigma (\mbox{\AA})$\\ 
\hline 
Na-O & 0.56014 & 2.720\\ 
Na-H & 0.56014 & 1.310\\ Cl-O & 1.50575 & 3.550\\ Cl-H &
1.50575 & 2.140\\ Na-Na & 0.11913 & 2.443\\ Cl-Cl & 0.97906 & 3.487\\
Na-Cl & 0.35260 & 2.796\\
\end{tabular}
\end{ruledtabular}
\end{center}
\label{tab:tab1}
\end{table}

Periodic boundary conditions are applied. 
The cutoff radius is fixed at $9.0$~\AA. 
Usually cutoff radius is fixed in simulations between $8$ and 
$10$~\AA~\cite{vega}.
Long range electrostatic interactions are taken into account by the
Ewald summation method with convergence parameter $\alpha$
set to $6.4/L$, where $L$ is the edge of the cubic
simulation box. 

The systems are equilibrated by controlling the temperature with the
Berendsen thermostat.~\cite{berendsen} Production runs are done in the 
$NVE$ ensemble. The integration timestep used is 1 fs.

For both systems we studied the densities $\rho= 1.125, 1.1, 1.05,
1.025, 0.98, 0.95, 0.90,  0.87, 0.85, 0.80\, g/cm^3$. 
For each density, a starting  configuration is produced
distributing the particles on a face centered cubic lattice, with
random orientation of water molecules. The crystal is then melted at
$T=500$~K and the temperature is stepwise reduced during the
equilibration runs. The lowest temperature investigated is $T=190$~K.
Equilibration runs become very long for
the lowest temperatures investigated. 
Each equilibration run is followed by a production run in which the
thermodynamic averages are calculated. Production runs
are always done with the same length of the equilibration runs. 
The longest equilibration and production runs last up to 10 ns each.
 
The simulations are 
carried out using the {\footnotesize DL\_~POLY} package.~\cite{dlpoly} 
The pressures extracted are calculated with the virial equation~\cite{allen}.

The choice of the force field is very important
in the case of ionic aqueous solutions~\cite{patra,cheatham}
since for example in KCl recent studies have evidenciated possible 
problems that are water model independent~\cite{kclpot}.
However the NaCl behavior seems to show an weaker 
dependence on the specific force field since it shows a 
lower tendency to form clusters~\cite{kclpot}. 
For the $c=2.10\, mol/kg$ solution and $\rho=1.1$ $g/cm^3$ at $T=300$~K
we have an internal energy value of $-62.81$ $kJ/mol$. This value can be compared 
with a simular value of $-69.57$ $kJ/mol$ 
obtained for a $c=2.35\, mol/kg$ solution and $\rho=1.093\, g/cm^3$ at $T=300$~K 
for a ionic potential with SPC flexible water potential~\cite{lyu}.
In order to stringently test the robustness of our potential 
we have run simulations along a isochore with a recent 
ionic potential by Jensen and Jorgensen~\cite{JJ} tailored for TIP4P water. 
In Fig.~\ref{fig:1} we show, for the two different concentrations, 
the isochore $\rho=1.025$ $g/cm^3$ calculated with both potentials.
We can see that the two potentials produce similar results.

\begin{figure}[t!] \centerline{\psfig{file=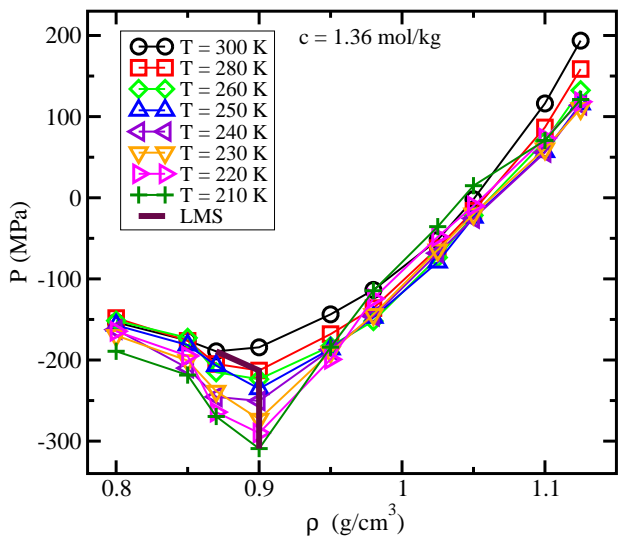,
width=0.5\textwidth}}
\caption{Isotherms in the range $210\,\mbox{K} \leq T \leq
300\,\mbox{K}$ and LMS line of $c=1.36\, mol/kg$ NaCl(aq) in the
$P-\rho$ plane (color online).}
\label{fig:3}
\end{figure}

\begin{figure}[t!] \centerline{\psfig{file=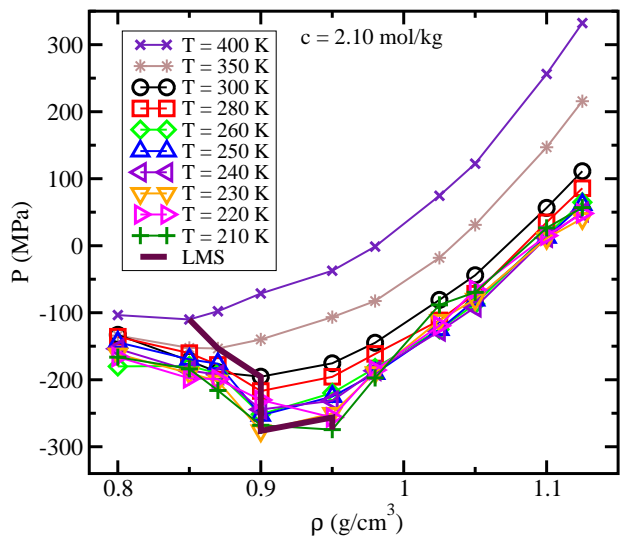,
width=0.5\textwidth}}
\caption{Isotherms in the range $210\,\mbox{K} \leq T \leq
400\,\mbox{K}$ and LMS line of $c=2.10\, mol/kg$ NaCl(aq) in the
$P-\rho$ plane (color online).}
\label{fig:4}
\end{figure}

We have also conducted a test to verify that our
data do not depend significatively on the size of the box.
Results are reported in Fig.~\ref{fig:2}.
For $c=1.36 \, mol/kg$  we compare the $\rho=0.98$  $g/cm^3$ isochore
calculated for 244 water molecules and 6 ion pairs, and
for 488 water molecules and 12 ion pairs. 
The simulation box 
of these systems is $L=20.037,\,25.2468$~\AA respectively.  
For $c=2.10\, mol/kg$ we compare the $\rho=0.98$  $g/cm^3$ isochore
calculated for 238 water molecules and 9 ion pairs,
for 476 water molecules and 18 ion pairs and
for 952 water molecules and 36 ion pairs. 
The simulation box 
of these systems is $L=20.132,\,25.365,\,31.958$~\AA respectively.  
We note that the curves corresponding 
to the same concentrations are very similar and that their minimum
does not show any significant shift.

\section{Thermodynamic Results}\label{thermo}

\begin{figure}[t!] \centerline{\psfig{file=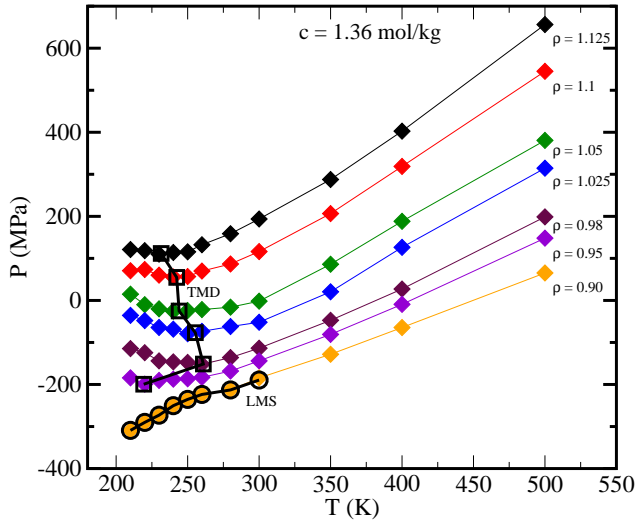,
width=0.5\textwidth}}
\caption{Isochores in the range $0.90\,g/cm^3 \leq \rho \leq
1.125\,g/cm^3$, TMD and LMS lines of $c=1.36\, mol/kg$ NaCl(aq) in the
$P-T$ plane (color online).}
\label{fig:5}
\end{figure}

The simulated thermodynamic state points have been reported in the
$P-\rho$ (isotherms) plane and in the $P-T$ (isochores) plane. The
analysis of those planes allows  the determination of the LMS line and
TMD line, respectively. Both curves can be derived using
thermodynamic relations. 

By considering the isothermal compressibility

\begin{equation} K_T=-\frac{1}{V}\left(\frac{\partial V}{\partial
P}\right)_T=\frac{1}{\rho}\left(\frac{\partial \rho}{\partial P}
\right)_T
\end{equation} 

the LMS line
is defined by the locus of the points for 
which $K_T$ diverges.
The line that joins the minima of the
isotherms corresponds to the LMS line. The TMD line is defined as
the locus of the points  where the coefficient of thermal expansion
$\alpha_P$ is zero.

\begin{equation} \alpha_P=\frac{1}{V}\left(\frac{\partial V}{\partial
T}\right)_P=-\frac{1}{\rho}\left(\frac{\partial \rho}{\partial T}
\right)_P=K_T\left(\frac{\partial P}{\partial T}\right)_\rho\, 
\end{equation} 

Therefore the line joining the minima of the isochores
yields the TMD line.

In Fig.~\ref{fig:3} and Fig.~\ref{fig:4} we report the isotherms
of the two solutions  in the $P-\rho$ plane 
as given by our simulations.  
In both cases we display only the curves
that show minima and thus contribute to the calculation of the LMS
line.  

\begin{figure}[t!] \centerline{\psfig{file=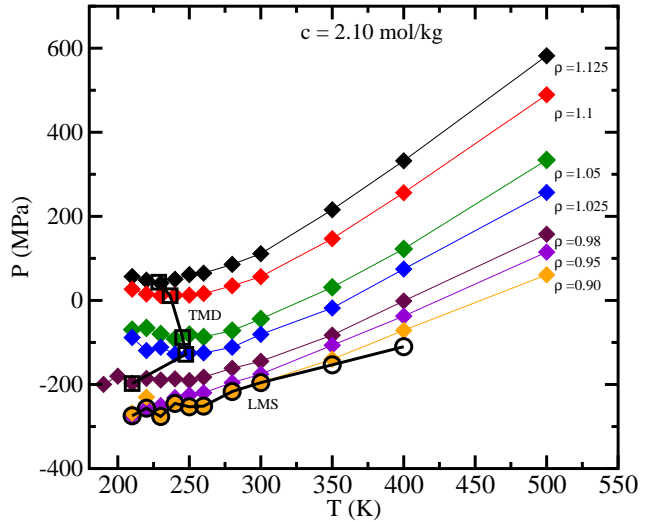,
width=0.5\textwidth}}
\caption{Isochores in the range $0.90\,g/cm^3 \leq \rho \leq
1.125\,g/cm^3$, TMD and LMS lines of $c=2.10\, mol/kg$ NaCl(aq) in the
$P-T$ plane (color online).}
\label{fig:6}
\end{figure}

Fig.~\ref{fig:3} refers to the $c=1.36\, mol/kg$ solution.
We show the isotherms in the range $210\,\mbox{K}
\leq T \leq 300\,\mbox{K}$ and the 
corresponding LMS line. 
$T=300$~K is the highest temperature isotherm which shows a minimum.
The LMS line starts at $\rho=0.87\, g/cm^3$ at $T=300$~K and shifts to 
$\rho=0.90\, g/cm^3$  for all the lower temperature curves.  
The lowest temperature isotherms, $T=220$~K and $T=210$~K, show
inflections that cross the higher temperature isotherms
for densities in the range 
$0.98\,g/cm^3 \leq \rho \leq 1.05\,g/cm^3$. 

Fig.~\ref{fig:4} refers to the $c=2.10\, mol/kg$ solution. 
We report the isotherms in the range
$210\,\mbox{K} \leq T \leq 400\,\mbox{K}$ and the 
corresponding LMS line.
At this concentration, minima of the
isotherms can be found up to the $T=400$~K isotherm. The LMS line 
gradually shifts
toward higher densities upon decreasing the temperature. In this case
the only isotherm showing an inflection is the one at $T=210$~K. 
This inflection spans the density range
$1.025\,g/cm^3 \leq \rho \leq 1.05\,g/cm^3$. 

Upon comparing the isotherms planes of the two solutions we note
that, at high densities, the isotherms of the higher concentration
solution are shifted by about 50 MPa toward lower pressures, with
respect to the isotherms of the lower concentration solution. This
shift decreases at densities close to the minima of the
isotherms and it almost disappears for very low densities. This
behavior of the isotherms is analogous to what found in the comparison
between bulk water and the $c=0.67\, mol/kg$ solution.~\cite{mio} 
A similar pressure shift can be seen also when comparing the $c=1.36\,
mol/kg$ and the $c=0.67\, mol/kg$ solutions (not shown). 
Therefore upon increasing ions concentration the isotherms 
progressively shift toward lower pressures. 

For the $c=1.36\, mol/kg$ solution the LMS line in the isotherms plane
(Fig.~\ref{fig:3}) is monotonic as already found for bulk
water~\cite{Gene, Spinodal, Mossa}, confined water~\cite{gallorovere}
and $c=0.67\, mol/kg$ NaCl(aq).~\cite{mio} In the $c=2.10\, mol/kg$
solution (Fig.~\ref{fig:4}) it does not decrease
on going from the $T=230$~K to the $T=220$~K isotherm. 

An important feature of the isotherms planes of the two solutions
is the presence of inflections of the low temperature
isotherms. It has been previously shown for bulk water that those inflections in
the isotherms are a signature of the approach of the systems to
liquid-liquid (LL) coexistence.~\cite{Gene, Spinodal,Mossa} As already
noted for the $c=0.67\, mol/kg$ solution~\cite{mio}, this behavior
is maintained in the NaCl(aq). Therefore we can infer that the HDL/LDL coexistence,
possibly terminating in a second critical point, is present in the NaCl(aq)
solutions, at least up to $c=2.10\, mol/kg$ concentration. Nonetheless the shrinkage
of the density range of inflections in the isotherms seems to indicate a gradual disappearance
of the coexistence upon increasing salt content.

These findings are
consistent with what found by Archer and Carter~\cite{Archer} in
their experimental paper. They found that the anomalous behavior of
supercooled water, and in particular the divergence of isobaric
specific heat and the existence of a TMD line are maintained in
NaCl(aq) up to concentrations of about $2\, mol/kg$. Thus, in the
framework of the second critical point scenario, it could be proposed
that those anomalies are a consequence of a second critical point in
the NaCl(aq) system, shifted toward lower pressures
with respect to bulk water.

\begin{figure}[t!] \centerline{\psfig{file=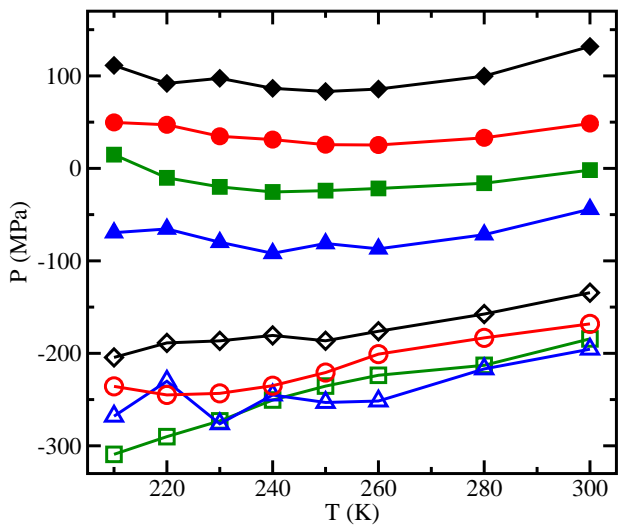,
width=0.5\textwidth}}
\caption{Isochores in the temperature range $210\,\mbox{K} \leq T \leq
300\,\mbox{K}$, starting from the top, for bulk water (diamonds), $c=0.67\, mol/kg$ (circles),
$c=1.36\, mol/kg$ (squares) and $c=2.10\, mol/kg$ (triangles) solutions at
densities  $\rho=$~1.05 (filled symbols) \mbox{and} 0.90 (unfilled symbols) $g/cm^3$ (color online).}
\label{fig:7}
\end{figure}

\begin{figure}[t!] \centerline{\psfig{file=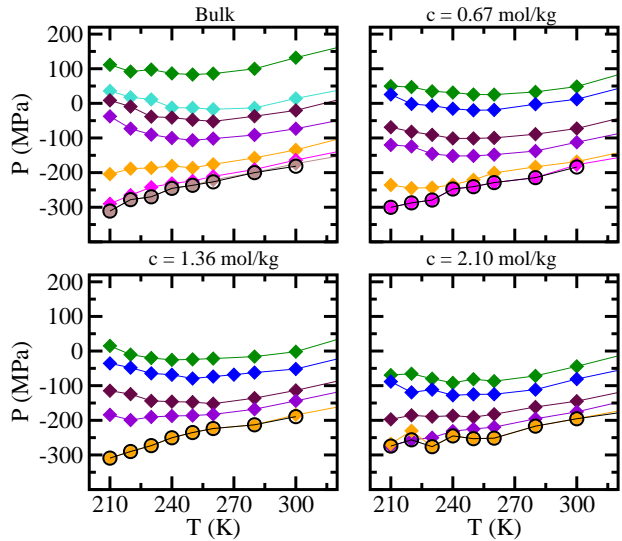,
width=0.5\textwidth}}
\caption{Isochores and LMS lines (open circles) in the $P-T$ plane for
 bulk water~\cite{mio} (top left panel), $c=0.67\, mol/kg$ NaCl(aq)~\cite{mio} 
(top right panel), $c=1.36\, mol/kg$ NaCl(aq) (bottom left panel) and $c=2.10\,
 mol/kg$ NaCl(aq) (bottom right panel), in the temperature range $210\,\mbox{K}
 \leq T \leq 300\,\mbox{K}$ and, starting from the
top, for densities $\rho=1.05,1.00,0.98.0.95,0.90,0.87,0.85\, g/cm^3$ for bulk water,
and for densities $\rho=1.05,1.025,0.98.0.95,0.90 \, g/cm^3$ for NaCl(aq) solutions 
(also $\rho = 0.87 \, g/cm^3$ only for the $c=0.67\, mol/kg$ solution), (color online).}
\label{fig:8}
\end{figure}

In Fig.~\ref{fig:5} and Fig.~\ref{fig:6} the isochores planes of the
two solutions are reported. The isochores are presented along
with the LMS lines and the TMD lines. For both systems the
isochores lying below the LMS line are not reported.  The minima are obtained by 
fitting the isochores with fourth degree polynomial functions. 

In Fig.~\ref{fig:5} we show the isochores in the range $0.90\,g/cm^3
\leq \rho \leq 1.125\,g/cm^3$, the LMS line and the TMD line for the
$c=1.36\, mol/kg$ solution. The range of temperatures spanned is
$210\,\mbox{K} \leq T \leq 500\,\mbox{K}$. All the isochores above
$\rho=0.90\, g/cm^3$ display a minimum, while the
$\rho=0.90\, g/cm^3$ isochore is almost completely coincident with the LMS
line. Such LMS line is entirely in the region of negative pressure and
it is nonre-entrant down to the lowest temperature we simulated. This behavior
has been already found in bulk water~\cite{Poole,Gene, Spinodal, Mossa, hpss, Netz}, confined
water~\cite{gallorovere} and $c=0.67\, mol/kg$ NaCl(aq).~\cite{mio}

In Fig.~\ref{fig:6} the isochores in the range $0.90\,g/cm^3 \leq \rho
\leq 1.125\,g/cm^3$, the LMS line and the TMD line for the $c=2.10\,
mol/kg$ solution are displayed. Also in this case the range of
temperatures spanned is $210\,\mbox{K} \leq T \leq 500\,\mbox{K}$. The
isochores above the $\rho=0.95\, g/cm^3$ show a
minimum. At this concentration 
some oscillations can be found in the LMS line
at low temperatures. 
This line approximately follows 
the trend found for the $\rho=0.90\, g/cm^3$ isochore. 

The comparison of the two isochores planes shows that also the
isochores of the $c=2.10\, mol/kg$ solution are shifted toward lower
pressures by roughly 50 MPa, with respect to the $c=1.36\, mol/kg$
solution, as already noted for the isotherms of the systems. This
pressure shift decreases at low densities and at $\rho=0.90\,
g/cm^3$ it almost vanishes.

In order to have a direct comparison of the two aqueous solutions studied here
with both bulk water
and the $c=0.67\, mol/kg$ solution, we report in Fig.~\ref{fig:7} the
isochores for bulk water, $c=0.67\, mol/kg$, $c=1.36\,
mol/kg$ and $c=2.10\, mol/kg$ solutions in the temperature range  $210\,\mbox{K} \leq T \leq
300\,\mbox{K}$ and for densities $\rho=0.90\,g/cm^3$ and
$\rho=1.05\,g/cm^3$. In this picture it is more evident that
the increase in concentration leads to a downward
pressure shift of the corresponding isochores. This shift is quite
consistent at high densities while it tends to reduce at low densities,
close to the LMS line. 
To best show the overall packing of the isochores 
upon increasing salt content in Fig.~\ref{fig:8} we show the blow-up 
of the $P-T$ plane in the temperature range $210\,\mbox{K} \leq T \leq 300\,\mbox{K}$ for all systems
in the same density range.
On going from bulk water to the $c=2.10\, mol/kg$ solution we observe a decrease by
about 50\% of the spanned range of pressures.

It has been previously observed in the literature that at constant pressure,
ions have an effect on the structure of liquid water equivalent to the
application of an external pressure.~\cite{leberman,botti,
mancinelli,pasch,pasch2} This effect is consistent with our findings
since we see that the bulk isochores at a certain density coincide  
with higher density isochores of the aqueous solution.

\begin{figure}[t] \centerline{\psfig{file=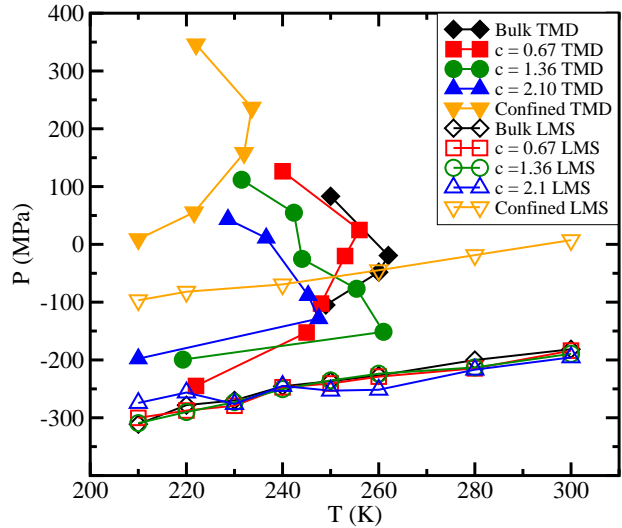,
width=0.5\textwidth}}
\caption{TMD and LMS lines in the $P-T$ plane for the $c=1.36\,
mol/kg$ and the $c=2.10\, mol/kg$ solutions, the $c=0.67\, mol/kg$
solution and bulk water studied in~\onlinecite{mio} and the hydrophobic
confinement system studied in~\onlinecite{gallorovere} (color online).}
\label{fig:9}
\end{figure}

In Fig.~\ref{fig:9} we show the comparison of TMD and LMS lines in the
$P-T$ plane among different systems:
the two solutions studied in the present paper, 
the $c=0.67\, mol/kg$ NaCl(aq) and 
bulk water~\cite{mio}
and TIP4P water confined in a hydrophobic environment of
soft spheres.~\cite{gallorovere} 
The temperature range spanned here is $210\,\mbox{K}
\leq T \leq 300\,\mbox{K}$. We note that the
LMS line appears not to be significantly affected
by the presence of ions. For the $c=0.67\, mol/kg$ and the  $c=1.36\,
mol/kg$ solutions the LMS line is almost identical to the bulk LMS. Only
at the highest concentration, $c=2.10\, mol/kg$, it shows minor
differences. 
In the case of water in hydrophobic matrix, which behaves similarly to
a solution of small apolar solutes~\cite{gallorovere},
the LMS line is similar in shape to that of bulk water and NaCl(aq) 
solutions but presents a significant shift, circa 200 MPa,
in the direction of higher pressures due to excluded volume effects
caused by the strong solute solvent repulsive strength.

At variance with the LMS line, 
the TMD line is markedly influenced by the presence of
ions. Upon increasing concentration, in fact, it shifts to lower
temperatures and pressures. Moreover in all our solutions the TMD line
extends to lower densities with respect to bulk water. Also the shape
is modified by the ions and this effect appears to be
influenced by concentration. For the $c=0.67\, mol/kg$ solution the
TMD line is much broader than for bulk water. For the $c=1.36\,
mol/kg$ and the $c=2.10\, mol/kg$ solutions, the TMD narrows,
remaining broader than that of bulk water.  In the case of the
hydrophobic solute, the TMD line is broadened with respect to
bulk water in a way similar to the case of $c=0.67\, mol/kg$ NaCl(aq),
but it is shifted by roughly 200 MPa (upwards) in pressure and 40~K
(to the left) in temperature with respect to the bulk.
A similar shift in temperature was found also by Kumar \emph{et al.} in 
TIP5P water confined between 
hydrophobic plates.~\cite{Kumar}

\begin{figure}[t] \centerline{\psfig{file=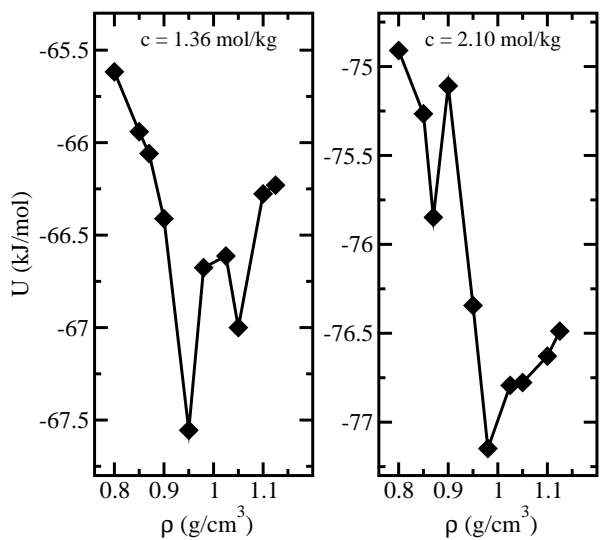,
width=0.5\textwidth}}
\caption{Potential energy per molecule at $T=210$~K, as a function of
density for the $c=1.36\, mol/kg$ solution (left panel) and  for the
$c=2.10\, mol/kg$ solution (right panel).}
\label{fig:10}
\end{figure}

Now we further inquire on the possibility of a LL coexistence in our solutions.
In Fig.~\ref{fig:10} we show the potential (or configurational) energy
per molecule of the two solutions, $U$, at $T=210$~K, as a function of the
density. For both solutions, $U$ shows two minima. For the $c=1.36\,
mol/kg$ solution, these minima are in correspondence with densities $\rho=0.95\,
g/cm^3$ and $\rho=1.05\, g/cm^3$. For the $c=2.10\, mol/kg$ solution,
minima can be found in correspondence with densities $\rho=0.87\,
g/cm^3$ and $\rho=0.98\, g/cm^3$.  The existence of minima in the
potential energy as a function of the density, at low temperatures,
can be related to the presence of LDL/HDL coexistence, as shown by
Kumar \emph{et al.}~\cite{Kumar} for water in hydrophobic 
confinement at $T=220$~K. Therefore the presence of two minima
in the potential energy, at low temperature, in our solutions
confirms the indication of the existence of a LL coexistence inferred
from the inflections of low temperatures isotherms that we observed. 

We can thus infer that, up to the highest concentration we
studied, the anomalous behavior of supercooled water is maintained in
the NaCl(aq) solutions. The presence of the ions 
does not seem to hinder the mechanism of emergence of this anomalous 
behavior. On the other hand the shift and the packing in the thermodynamic plane 
and the gradual weakening of the inflection in the isotherms, upon
increasing salt content, are signatures
of a progressive disappearance of the anomalies for higher concentrations
of ions.

\section{Conclusions}\label{conclu}

We performed MD simulations on two sodium chloride aqueous solutions in TIP4P
water, with concentrations $c=1.36\, mol/kg$ and $c=2.10\, mol/kg$, extending our
previous work on bulk water and $c=0.67\, mol/kg$ NaCl(aq).~\cite{mio}
Using the simulated state points, we drew the isotherms and the isochores planes of
the systems. The analysis of those planes allowed the determination of the LMS line and
of the TMD line, respectively.

By comparing the results obtained for the solutions studied here and 
in previous work, we can
observe that the presence of minima in the isotherms, determining the presence of a 
LMS line, is preserved in all the NaCl(aq) solutions analyzed. 
Importantly, also the inflections 
for low temperatures isotherms, at high densities, are maintained in the solutions.
Nonetheless, they become less pronounced upon increasing concentration. As previously 
shown for bulk water~\cite{Gene, Spinodal,Mossa}, those inflections signal the presence
of phase coexistence between LDL and HDL in the systems. 
Being the LL coexistence still present up to the highest concentration we studied,
we can hypothesize that whether a second critical
point exists for bulk water, it is preserved in the NaCl(aq) solution. 
Also the existence of a TMD line, determined by the presence 
of minima in the isochores plane, is maintained in the NaCl(aq) solutions.
Experimental results have shown that the TMD 
at ambient pressure is still present for a $c=1.49 \, mol/kg$ solution and 
disappears for $c=2.33\, mol/kg$.~\cite{Archer}  In our results we note that
the minima of the isochores become less pronounced for the $c=2.10\, mol/kg$
solution and correspondingly the TMD narrows.

Although the overall thermodynamic behavior of the NaCl(aq) solution is similar to that
of bulk water, some differences can be noted. 
In fact, for densities high with respect to the
LMS line, both the isotherms and the isochores shift to lower pressures upon increasing 
concentration.
This behavior results in a significant packing of the curves, 
as it is particularly evident in the isochores plane. The difference 
of pressure between the highest density isochore and the LMS 
line is in fact much broader in bulk water than in solutions 
and it is reduced upon increasing concentration. 
The LMS line is not influenced by the presence of ions. In fact both the position 
in the $P-T$  plane and the shape remains similar to bulk water, 
with some minor differences appearing for the highest concentration studied. 
The TMD line is instead modified in shape and shifted 
toward lower pressures and temperatures upon increasing concentration. 

The results for the LMS line and TMD line were also compared to those for water 
confined in a hydrophobic environment 
of soft spheres.~\cite{gallorovere} For this system the LMS line is similar 
in shape to that of 
bulk water but it is shifted to higher pressures. The TMD line is instead similar to that 
of $c=0.67\, mol/kg$ NaCl(aq) in shape but it is shifted to higher pressures and 
lower temperatures.

The existence of two minima in the curves of the potential energy as
function of density indicates that LL coexistence is present in the 
the NaCl(aq) solutions studied here, 
in analogy with what found for water confined between two hydrophobic 
plates.~\cite{Kumar} This result strengthen the hypothesis 
of a LL coexistence deduced from the observation of inflections 
in the low temperature isotherms. 

We note on passing that signatures of LL coexistence have
been experimentally found in LiCl(aq) solutions.~\cite{Mishima2} 

We have shown that various anomalous features 
of supercooled water are preserved 
in aqueous solutions of sodium chloride up to the highest concentration 
investigated.
Interestingly the TMD can be even followed 
down to lower densities with respect to 
bulk. From the thermodynamic features 
investigated we hypothesize that the LL critical point would be 
slightly shifted in temperature and
more markedly shifted in pressure with respect to bulk water.
Therefore an experimental observation of the LL critical point in
aqueous solutions, for which crystallization is more easily avoided 
than in the bulk phase~\cite{Angell2}, could be possible.

\section*{ACKNOWLEDGMENTS}
The authors gratefully acknowledge the computational support of the
Democritos National Simulation Center, 
at SISSA, Trieste and of the Roma Tre INFN-GRID.

\end{document}